\documentclass[twocolumn,10pt,showpacs,notitlepage,floatfix,superscriptaddress,amsmath,amssymb,
nofootinbib,floatfix
]{revtex4-2}

\usepackage{algpseudocode}
\usepackage{algorithm}
\usepackage[normalem]{ulem}
\usepackage{kpfonts}
\usepackage{graphicx}
\usepackage{amssymb}
\usepackage{amsmath}
\usepackage{amsthm}
\usepackage{amsfonts}
\usepackage{mathtools}
\usepackage{xcolor}
\usepackage{hyperref}
\usepackage{tikz}

\usetikzlibrary{cd}

\newtheorem*{lem*}{Lemma}
\newtheorem*{thm*}{Theorem}

\newcommand{\eq}[1]{\hyperref[eq:#1]{Eq.~(\ref*{eq:#1})}}
\renewcommand{\sec}[1]{\hyperref[sec:#1]{Section~\ref*{sec:#1}}}
\newcommand{\secsm}[1]{\hyperref[sec:#1]{Sec.~\ref*{sec:#1}}}
\newcommand{\app}[1]{\hyperref[app:#1]{Appendix~\ref*{app:#1}}}
\newcommand{\theo}[1]{\hyperref[thm:#1]{Theorem~\ref*{thm:#1}}}
\newcommand{\algo}[1]{\hyperref[alg:#1]{Algorithm~\ref*{alg:#1}}}
\newcommand{\lemm}[1]{\hyperref[lem:#1]{Lemma~\ref*{lem:#1}}}
\newcommand{\defn}[1]{\hyperref[defn:#1]{Definition~\ref*{defn:#1}}}
\newcommand{\corr}[1]{\hyperref[cor:#1]{Corollary~\ref*{cor:#1}}}
\newcommand{\condition}[1]{\hyperref[cond:#1]{Condition~\ref*{cond:#1}}}
\newcommand{\fig}[1]{\hyperref[fig:#1]{Fig.~\ref*{fig:#1}}}
\newcommand{\tab}[1]{\hyperref[tab:#1]{Table~\ref*{tab:#1}}}
\newcommand{\tabsm}[1]{\hyperref[tab:#1]{Tab.~\ref*{tab:#1}}}
\newcommand{\propos}[1]{\hyperref[prop:#1]{Proposition~\ref*{prop:#1}}}
\newcommand{\propsm}[1]{\hyperref[prop:#1]{Prop.~\ref*{prop:#1}}}
\newcommand{\rema}[1]{\hyperref[rem:#1]{Remark~\ref*{rem:#1}}}

\newcommand{\width}{W}
\newcommand{\offset}{F}
\newcommand{\ztwo}{\mathbb{F}_2}

\newcommand{\transpose}{\mathsf{T}}
\newcommand{\decodewin}{\mathcal{D}_{\textrm{win}}}
\newcommand{\Hwin}{H_{\textrm{win}}}

\newcommand{\idealdecoder}{\mathcal{D}_{\textrm{ideal}}}

\begin{document}

\title{Improved Noisy Syndrome Decoding of Quantum LDPC Codes with Sliding Window}

 \author{Shilin Huang}
 \affiliation{Department of Applied Physics, Yale University, New Haven, CT 06511}
 \affiliation{Yale Quantum Institute, Yale University, New Haven, CT 06520}
 \author{Shruti Puri}
 \affiliation{Department of Applied Physics, Yale University, New Haven, CT 06511}
 \affiliation{Yale Quantum Institute, Yale University, New Haven, CT 06520}

\begin{abstract}
Quantum error correction (QEC) with single-shot decoding enables reduction of errors after every single round of noisy stabilizer measurement, easing the time-overhead requirements for fault tolerance. Notably, several classes of quantum low-density-parity-check (qLDPC) codes are known which facilitate single-shot decoding, potentially giving them an additional overhead advantage.  
However, the perceived advantage of single-shot decoding is limited because it can significantly degrade the effective code distance. 
This degradation may be compensated for by using a much larger code size to achieve the desired target logical error rate, at the cost of increasing the amount of syndrome information to be processed, as well as, increasing complexity of logical operations. Alternatively, in this work we study {\it sliding-window} decoding, which corrects errors from previous syndrome measurement rounds while leaving the most recent errors for future correction. We observe that sliding-window decoding significantly improves the logical memory lifetime and hence the effective distance compared to single-shot decoding on hypergraph-product codes and lifted-product codes. Remarkably, we find that this improvement may not cost a larger decoding complexity. Thus, the sliding-window strategy can be more desirable for fast and accurate decoding for fault-tolerant quantum computing with qLDPC codes.

\end{abstract}

\maketitle

\section{Introduction}
Quantum low-density-parity-check (qLDPC) codes with a constant encoding rate have emerged as a promising approach for reducing the overhead of fault-tolerant quantum computation~\cite{gottesman_fault-tolerant_2014, fawzi_constant_2020,breuckmann_quantum_2021}. 
In recent years, several breakthroughs were made on improving the asymptotic distance of qLDPC codes~\cite{hastings_fiber_2021,panteleev_quantum_2022,breuckmann_balanced_2021}.
In particular, Panteleev and Kalachev first proved that qLDPC codes with a constant encoding rate can achieve a code distance asymptotically linear in the number of physical qubits~\cite{panteleev_asymptotically_2022}. Since then, alternative constructions of good qLDPC code have been proposed~\cite{dinur_good_2023}. 
Numerical studies~\cite{tremblay_constant-overhead_2022,bravyi_high-threshold_2023,xu_constant-overhead_2023,higgott_subsystem_2021,higgott_constructions_2023} suggest that qLDPC codes using considerably fewer physical qubits can still achieve the same amount of error suppression as with the planar surface codes~\cite{dennis_topological_2002}, the leading approach for practical quantum error correction (QEC)~\cite{fowler_surface_2012}. These promising results, combined with the rapid technological progress in realizing long-range connectivity~\cite{bluvstein_quantum_2022,pino_demonstration_2021,magnard_microwave_2020} is making qLDPC codes a promising candidate for practical QEC.

In order to achieve constant-rate with qLDPC codes in practice, it is necessary to have efficient decoding algorithms to process the syndromes accumulated from stabilizer measurements and identify the errors~\cite{leverrier_quantum_2015,leverrier_decoding_2023,delfosse_toward_2022,panteleev_degenerate_2021,roffe_decoding_2020,dinur_good_2023,gu_efficient_2023}. 
Remarkably, many qLDPC code families have enough in-built redundancy to allow for reliable identification of errors even if some of the stabilizers are measured incorrectly~\cite{fawzi_constant_2020,panteleev_degenerate_2021,bombin_single-shot_2015,kubica_single-shot_2022,gu_single-shot_2023}. This feature implies
that a single round of syndrome measurement is sufficient for fault-tolerant error correction, making it possible to have {\it single-shot decoding}~\cite{bombin_single-shot_2015,campbell_theory_2019}. Many numerical simulations of qLDPC codes have observed high fault-tolerance thresholds with single-shot decoding~\cite{brown_fault-tolerant_2016,
kubica_single-shot_2022,grospellier_combining_2021,tremblay_constant-overhead_2022,higgott_improved_2023,xu_constant-overhead_2023,quintavalle_single-shot_2021,breuckmann_single_2022}. This is in stark contrast with the standard implementation of the planar surface code, which requires a number of stabilizer measurements scaling with the code size for fault tolerance~\cite{dennis_topological_2002}.

Despite its promise, single-shot decoding is sub-optimal as it often fails to preserve the code distance and degrades the scaling of the logical error rate in the sub-threshold regime. Consequently, a much larger code size becomes necessary to achieve a given target logical error rate, which is experimentally challenging in the near term. Moreover, this solution comes at the cost of increase in the space-time overhead for implementing logical quantum operations~\cite{gottesman_fault-tolerant_2014,cohen_low-overhead_2022}, making it undesirable even in the long term. Thus, it is necessary to find alternative efficient decoding strategies with minimal degradation of the code distance.


With this motivation, in this work we study the performance of \emph{sliding window} decoding for hypergraph-product codes~\cite{tillich_quantum_2014,leverrier_quantum_2015} and lifted-product codes~\cite{panteleev_quantum_2022},
which are qLDPC codes of substantial theoretical and practical interests. The sliding-window strategy and its variants have been extensively studied for decoding surface codes~\cite{dennis_topological_2002,huang_between_2021, skoric_parallel_2023,tan_scalable_2022,bombin_modular_2023,berent2023analog}, but a systematic study for qLDPC codes is lacking. Distinct from single-shot decoding, a sliding-window decoder uses syndromes across a window of multiple rounds of stabilizer measurements.
In order to maximize the lifetime of a qLDPC code memory, we find it crucial for the sliding window decoder to exclusively correct errors identified during the first few measurement rounds, while deferring the correction of the remaining errors to the subsequent error correction cycle. 
We observe that sliding window decoding significantly improves the expected lifetime of quantum memory under a phenomenological noise model where errors occur on data qubits and measurement bits but the stabilizer measurement circuit is assumed to be noise free. 
Importantly, we find that the sliding window decoder can achieve the same or longer memory lifetime as single-shot decoding without increasing the decoding complexity. Thus, we believe that the sliding-window strategy may be more suitable for fast and accurate decoding for fault-tolerant quantum computing.

This paper is organized as follows. \sec{ss_intro}
and~\sec{sw_intro} provide an overview of the single shot and sliding window decoding strategies respectively. 
We describe the numerical simulations and present the results in \sec{numerics}. Finally we conclude in \sec{outlook} by describing some open questions and future directions.

\section{Single-shot decoding}
\label{sec:ss_intro}

In this work, we use the notation $[[n,k,d]]$ to denote a quantum code that encodes $k$ logical qubits in $n$ physical qubits and has a distance $d$.
We focus on Calderbank-Shor-Steane (CSS) codes~\cite{calderbank_good_1996,steane_error_1996},
where $X$ errors (bit flips) and $Z$ errors (phase flips) can be decoded separately. 
Without loss of generality, we only consider decoding $X$ errors by measuring $Z$-type stabilizer generators of the code, which are represented by an $m\times n$ parity-check matrix $H$.
Given an $X$ error set represented by some binary vector $e \in \ztwo^n$, its syndrome should ideally be $\sigma_{\textrm{ideal}} := He \in \ztwo^m$.
Realistically, however, one might instead obtain a noisy syndrome $\sigma := \sigma_{\textrm{ideal}} + u$, where $u \in \ztwo^m$ represents the set of measurement errors.

A common strategy to filter out readout errors is to measure the syndrome repetitively~\cite{shor_fault-tolerant_1996,kitaev_quantum_1997}.
To perform faster error correction, however, one could consider single-shot decoding, 
which corrects the qubit errors immediately after a single round of noisy syndrome extraction~\cite{bombin_single-shot_2015}. Given a noisy syndrome $\sigma$, a single-shot decoder finds a qubit error set $\tilde{e}$ and a measurement error set $\tilde{u}$ satisfying $H\tilde{e}+\tilde{u} = \sigma$. 
The correction $\tilde{e}$ is certainly not guaranteed to be equivalent to $e$ up to $X$-type stabilizer elements. However, single-shot correction is still considered successful if the residual error $e+\tilde{e}$ is equivalent to an $X$ error set $r$ whose weight $|r|$ is bounded. 

Often  single-shot decoding is considered valid on a qLDPC code if a fault-tolerance threshold exists~\cite{campbell_theory_2019}.
To yield a threshold, it is oftentimes sufficient to have
$|r| \le \alpha |u|$, where $\alpha > 0$ is some constant independent of code size $n$.
However, the threshold value is not the only indicator of the performance of QEC.
Given a physical error rate $p$ far below threshold, the logical error rate $p_{\textrm{L}}$ scales as 
$p_{\textrm{L}} \propto p^{(d_{\textrm{eff}}+1)/2}$, where $d_{\textrm{eff}}$ is known as the \emph{effective distance}. The effective code distance is upper bounded by the code distance $d$ and determines the size of the code required to reach a target error rate.

If $\alpha \le 1$, we can have $d_{\textrm{eff}} = d$ since 
single-shot decoding does not introduce correlated qubit errors.
However, if $\alpha \gg 1$, decoding a noisy syndrome with $|u|$ errors can introduce correlated qubit errors of weight $\alpha|u| \gg |u|$. Consequently, in the worst case, $d_{\textrm{eff}}$ can be degraded from $d$ to $d/\alpha$. As a result, a significantly larger code block is required to compensate for degradation of distance and achieve a target logical error rate.
While the overall space overhead may not change using constant-rate qLDPC codes, 
having more logical qubits encoded into a single block
can significantly increase the space-time overhead of logical computation~\cite{gottesman_fault-tolerant_2014}. Moreover, a larger code  also increases the decoding complexity and latency as the number of syndrome bits that must be processed increases.

\section{Sliding window decoding}
\label{sec:sw_intro}
As an alternative to increasing the code block to compensate for decreased effective distance with single-shot decoding, we consider sliding-window decoding~\cite{dennis_topological_2002,skoric_parallel_2023,tan_scalable_2022}, which uses a larger set of syndromes accumulated over multiple rounds of stabilizer measurements. 

Sliding-window decoding was first proposed for planar surface codes by Dennis \emph{et.al.} with its original name \emph{overlapping recovery}~\cite{dennis_topological_2002}.
In this procedure, one first decodes over $\width$ rounds of syndrome measurements. A correction is applied to, or is committed to, the first $\offset$ rounds for some $\offset \le \width$, leaving the errors in the last $\width-\offset$ rounds to the following error correction cycles.
In every subsequent cycle, the decoding window of \emph{width} $\width$ slides over the time axis by the \emph{offset} $\offset$ and errors are committed to in $\offset$.
For convenience, we call a sliding window decoding scheme with width $\width$ and offset $\offset$ as $(\width,\offset)$-decoding.
As an example, single-shot decoding is a $(1,1)$-decoding.
When $\offset < \width$, the decoding windows will overlap each other and thus we say that $(\width,\offset)$-decoding is an \emph{overlapping} decoding.
We use the term \emph{non-overlapping} decoding to refer to the cases where $\offset = \width$.

The formal procedure of a $(\width,\offset)$-decoding is described as follows.
Suppose a CSS code block with $Z$-type parity-check matrix $H$ begins to experience errors from time $0$. 
For each positive integer time $t$, 
a set of qubit errors occur in the time interval $(t-1,t)$, represented by a binary vector $e_t \in \ztwo^n$.
Syndrome extraction is performed at time $t$, whose measurement outcome can experience a set of measurement errors, represented by $u_t \in \ztwo^m$.
The measured syndrome at time $t$, denoted by $\sigma_t$, is therefore $\sigma_t = H\left(\sum_{j=1}^w e_j\right) + u_t$.

To perform $(\width,\offset)$-decoding, we first specify the window decoder $\decodewin$ which takes the list of syndrome vectors $\left(\sigma_1, \ldots, \sigma_{\width}\right)$ as input, and estimates the error vectors $\tilde{e}_1 \ldots, \tilde{e}_{\width}$, $\tilde{u}_1 \ldots, \tilde{u}_{\width}$
such that
$H\left(\sum_{j=1}^t \tilde{e}_j\right) + \tilde{u}_t = \sigma_t$
for all $t = 1, \ldots, \width$.
After $\decodewin$ is executed, we 
correct all the qubit and measurement errors in time interval $(0,\offset]$ and keep the updated syndrome vectors in time interval $[\offset+1, \width]$ for the next error correction (EC) cycle.
Operationally, we apply an overall $X$ correction $\xi := \sum_{j=1}^{\offset} \tilde{e}_j$ on the qubits, then update each $\sigma_t$ 
($t \in [\offset+1, \width]$) as $\sigma_t' := \sigma_t + H\xi$. The syndrome vectors in time interval $(0,\offset]$ are no longer needed for future EC cycles and therefore discarded.
{In practice, instead of applying this correction physically, one can simply update the syndromes in all subsequent time steps.}

In the next EC cycle, we perform $\offset$  rounds of syndrome measurements 
at times $t = \width+1, \ldots, \width+\offset$
and read out a sequence of syndrome vectors $\sigma_{\width+1} \ldots, \sigma_{\offset+\width}$. 
The input for the window decoder $\decodewin$ will be the list
$\left(\sigma_{\offset+1}', \ldots, \sigma_{\width}', \sigma_{\width+1}, \ldots, \sigma_{\offset+\width}\right)$. 
If syndrome measurements are performed indefinitely, the spans of decoding window will be the following time intervals 
$$(0, \width],\ (\offset, \offset+\width],\ (2\offset, 2\offset+\width]\ldots,$$ 
which overlap with each other if $F<W$.

We are interested in how different choices of $(\width,\offset)$ affect the performance of qLDPC codes on which $(1,1)$-decoding is sufficient to reach a threshold. 
Ref.~\cite{xu_constant-overhead_2023} studied $(3,3)$-decoding on hypergraph-product codes~\cite{tillich_quantum_2014} and lifted-product codes~\cite{panteleev_quantum_2022} and reported that $(3,3)$-decoding can achieve lower logical error rates than $(1,1)$-decoding.
However, non-overlapping decoding with a larger width $\width$ can still degrade the effective distance.
To see this, consider a scenario where qubit and measurement errors are all occurring in the last round, while the previous $\width-1$ rounds are noiseless.
As one cannot discriminate these data and measurement errors without syndromes in the future, 
$(\width,\width)$-decoding will lead to the same correlated residual error pattern as $(1,1)$-decoding on the last syndrome. As we will see in the next section, overlapping decoding is necessary for improvement of decoding accuracy beyond the non-overlapping strategy.


\section{Numerical Simulations}
\label{sec:numerics}
\begin{figure*}
   \includegraphics[width=\textwidth]{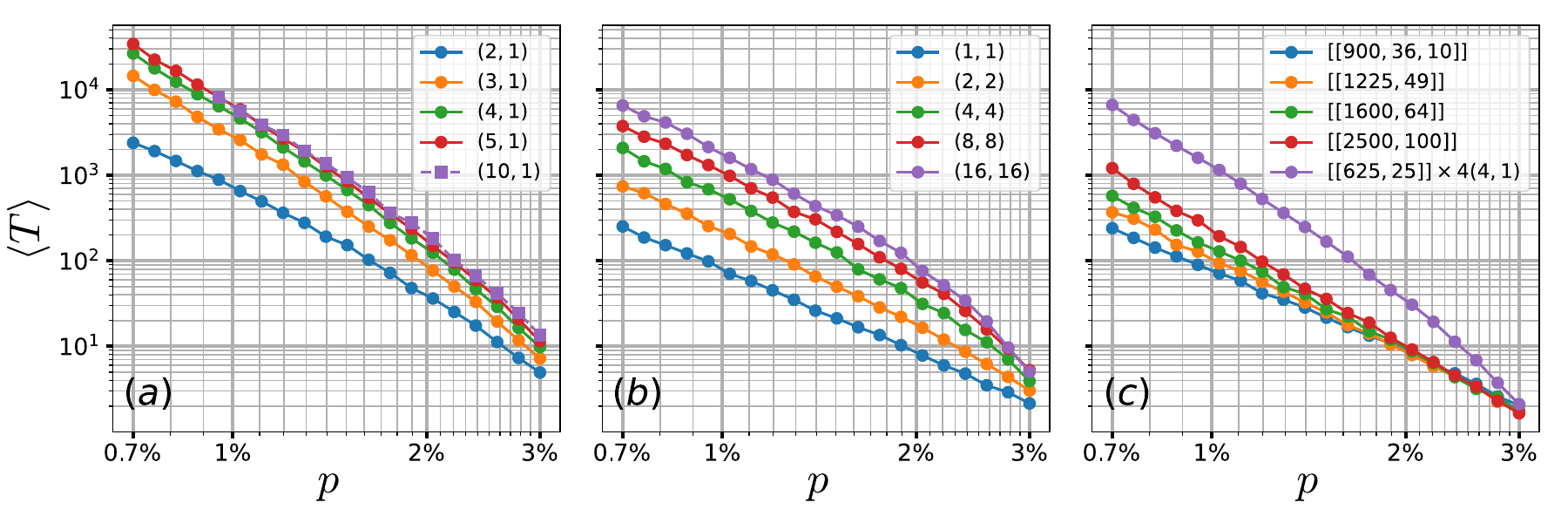} 
   \caption{{The expected logical memory lifetime of sliding-window $(\width,\offset)$-decoding 
   on hypergraph-product (HGP) codes of various sizes under phenomenological noise model. 
   $\width$ and $\offset$ are the width and offset of the decoding window respectively.
   The sampling errors are negligible.
   (a) Overlapping $(\width,1)$-decoding on $[[625,25,8]]$ code. (b) Non-overlapping $(\width,\width)$-decoding on $[[625,25,8]]$ code.
   (c) Comparison of single-shot $(1,1)$-decoding on larger HGP codes and $(4,1)$-decoding on 4 copies of $[[625,25,8]]$ code.}
   }\label{fig:hgp}
\end{figure*}

We numerically examine sliding window decoding with different width $\width$ and offset $\offset$ on various instances of qLDPC codes. 
Our noise model is  phenomenological:
for each round of syndrome extraction, an $X$ error is applied on each qubit independently with probability $p$ before the measurements, while each measurement outcome is flipped independently with probability $p$.

To benchmark the performance of $(\width,\offset)$-decoding on a chosen quantum code, we repeatedly apply noisy syndrome extractions and $(\width,\offset)$-decoding until a logical failure occurs. If a logical failure first occurs after $N$ EC cycles, the \emph{lifetime} of the quantum memory is defined as $T := (N-1) \times F$.
For a given code, we are interested in how the expected lifetime $\langle T \rangle$ scales with the physical error rate $p$. 

To determine whether a logical failure occurs after $N$ EC cycles, the simulator keeps track of the \emph{residual qubit error} $r := \sum_{i=1}^{N\offset} e_i + \sum_{j=1}^{N}\xi_j$ 
at time $N\offset$, where $\xi_j$ is the overall $X$ correction applied in $j$-th EC cycle. We say that a logical failure occurs if $r$ is not a correctable error \sout{set} even in the absence of measurement errors.
To determine whether $r$ is correctable or not, the simulator applies an \emph{ideal decoder} $\idealdecoder: \ztwo^m \rightarrow \ztwo^n$ which decodes the noiseless syndrome $Hr$ and returns $\tilde{r} := \idealdecoder(Hr)$, an estimation of $r$.
$r$ is not correctable if  $r+\tilde{r}$ is a non-trivial logical operator. 

A decoding algorithm is needed for implementing the decoders $\idealdecoder$ and $\decodewin$. Maximum-likelihood (ML) decoding, while being ideal for maximizing the lifetime,  is unfortunately hard to scale. 
A widely-used heuristic decoding algorithm for classical LDPC codes is \emph{belief propagation} (BP)~\cite{mackay_near_1997}. However, on qLDPC codes, BP often outputs an error inconsistent with the given syndrome due to the inherent degeneracy of the codes. These decoding failures can be significantly suppressed by post-processing the soft information provided by BP~\cite{panteleev_degenerate_2021,grospellier_combining_2021,higgott_improved_2023}. A notable post-processing algorithm is ordered-statistics decoding (OSD)~\cite{panteleev_degenerate_2021}, which has been tested on various qLDPC code instances 
for both noiseless and noisy syndrome measurements~\cite{panteleev_degenerate_2021,roffe_decoding_2020,quintavalle_single-shot_2021,higgott_improved_2023}.

In this work, for both $\idealdecoder$ and $\decodewin$ we use the open-source BP-OSD implementation developed by Roffe \emph{et.al.} \cite{roffe_decoding_2020}. Their implementation  uses the combination-sweep heuristic postprocessing strategy to find higher order solutions to BP-OSD for improved decoding. In our simulations, the maximum number of BP iterations is set to be the number of errors in the decoding problem and the combination-sweep parameter ($\lambda$) is set to $40$. 
We refer the reader to~\cite{panteleev_degenerate_2021} and ~\cite{roffe_decoding_2020} for a detailed discussion of BP-OSD decoder and the combination-sweep method.


The BP-OSD algorithm takes as input the set of observed syndromes and a matrix that maps errors to syndromes.
A sparse matrix is desirable otherwise the performance of BP can degrade.
For ideal decoder $\idealdecoder$, these elements are the syndrome of residual error $r$ and the parity-check matrix $H$. $H$ is sparse by definition of LDPC. For window decoder $\decodewin$, 
we let the input syndromes be $(\sigma_1, \sigma_2 - \sigma_1 \ldots, \sigma_\width - \sigma_{\width-1})$
instead of the original syndromes $(\sigma_1, \ldots, \sigma_\width)$.
The mapping
\begin{eqnarray*}
\left(e_1, \ldots, e_\width, u_1, \ldots, u_\width\right)
    \rightarrow 
    \left(\sigma_1, \sigma_2 - \sigma_1, \ldots, \sigma_\width - \sigma_{\width-1}\right)
\end{eqnarray*}
is a sparse parity-check matrix
    $\Hwin = \left[ I_\width \otimes H\middle| B \otimes I_m\right]$,
where $B$ is a $\width\times \width$ matrix such that $B_{ij} = 1$ if and only if $i=j$ or $i=j+1$. The sparsity of $\Hwin$ is promised by the sparsity of $H$ and $B$.


Our numerical study mainly focuses on hypergraph product (HGP) codes~\cite{tillich_quantum_2014}, the earliest  constant-rate qLDPC code family known to have a threshold with single-shot decoding~\cite{fawzi_constant_2020}. Nonetheless we will also show results of numerical simulations on the lifted product codes~\cite{panteleev_quantum_2022} which have been recently considered for implementation with atomic qubits~\cite{xu_constant-overhead_2023}.

\emph{Hypergraph Product Codes:}  We consider hypergraph 
product of a classical linear code with an $m_A\times n_A$ parity-check matrix $A$ and itself, denoted by $\textrm{HGP}(A)$. $\textrm{HGP}(A)$ has $X$ and $Z$ parity-check matrices of the following form:
\begin{equation*}
   H_X = \left[ A \otimes I_{n_A} \middle| I_{m_A} \otimes A^\transpose\right],\quad H_Z = \left[ I_{n_A} \otimes A \middle| A^\transpose \otimes I_{m_A}\right].
\end{equation*}
We further assume that $A$ is a randomly generated $(3,4)$-regular LDPC matrix. Here we say that a matrix is $(r,s)$-regular LDPC if each column of $A$ has exactly $r$ non-zero entries, while each row of $A$ has exactly $s$ non-zero entries.
It has been numerically verified that on this restricted HGP code family, single-shot decoding can achieve a threshold~\cite{grospellier_combining_2021}.

As an example, we perform sliding-window decoding
on the $[[625,25,8]]$ HGP code provided in Ref.~\cite{roffe_decoding_2020}.
The parity-check matrix $A$ of its base code are given in \app{HGP}. 
We estimate the expected memory lifetime $\langle T\rangle$ as a function of physical error rate $p$ using different widths $\width$ and offsets $\offset$.
Two decoding strategies are considered: (i) non-overlapping decoding, where $\offset = \width$ and (ii) overlapping decoding, with $\offset = 1$.

The results for overlapping and non-overlapping decoding are shown in \fig{hgp} (a) and (b) respectively. 
For overlapping decoding, we observe that the performance of $(5,1)$- and $(10,1)$-decoding is almost identical. We believe that it is unlikely that $\langle T\rangle$ will improve much more by further increasing $\width$.
For non-overlapping decoding, 
we observe that $\langle T\rangle$ improves more slowly as $\width$ increases. 
We expect that the expected lifetime of non-overlapping decoding becomes identical to overlapping decoding when $\width$ is sufficiently large.
Importantly, we find that the non-overlapping decoding strategy can be much worse than the overlapping strategy. 
As an example, $(16,16)$-decoding results in a lifetime that is nearly an order of magnitude smaller than the lifetime with $(3,1)$-decoding at $p=0.7\%$, which suggests that non-overlapping decoding does indeed introduce unwanted correlated residual qubit errors.

\fig{hgp}(c) shows the performance of single-shot decoding for  random HGP codes of various sizes.
The parity-check matrices of their base code is shown in \app{HGP}.
The $[[900,36,10]]$ codes is taken from Ref.~\cite{roffe_decoding_2020} and the $[[1225,49]]$, $[[1600,64]]$ and $[[2500,100]]$ codes are generated using the methods presented in Ref.~\cite{grospellier_combining_2021}.
We find that the expected lifetime of $[[625,25,8]]$ and $[[900,36,10]]$ codes are identical, which confirms that single-shot decoding can degrade effective code distance. 

Finally, a comparison between \fig{hgp}(a) and \fig{hgp}(c) also shows that the lower logical error rate with the overlapping window decoding does not come at the cost of larger decoding complexity. To see this, consider four copies of the $[[625,25,8]]$ code and a single block of the $[[2500,100]]$ code so that they have the same number of logical qubits. From \fig{hgp}(c) we see that the lifetime of four copies of the $(4,1)$-decoded $[[625,25,8]]$ code is larger than that for the $[[2500,100]]$ code with single-shot decoding. Next consider the number of syndrome bits that must be processed or the decoding volume $V=\width(n-k)/{2}$ as a measure of decoding complexity. This volume is $12\width k$ for the $[[25k,k]]$ family of HGP codes considered here. For the overlapping $(4,1)$-decoding on $[[625,25,8]]$ code, the decoding volume is $V=1200$ which is the same as the single-shot decoding volume for the $[[2500,100]]$ code even though it  yields longer normalized lifetimes. Note that unlike the lifetime, we do not compare the decoding volume normalized for the number of logical qubits. This is because only syndromes from with the same block are processed together but syndromes from different blocks can be processed independently. 

\begin{figure}
   \includegraphics[width=\linewidth]{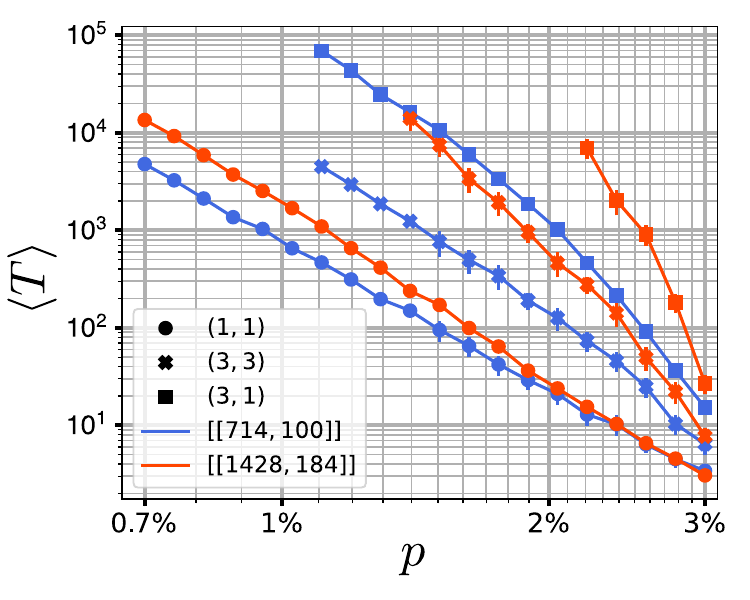} 
   \caption{{The expected logical memory lifetime of single-shot $(1,1)$-, non-overlapping $(3,3)$- and overlapping $(3,1)$-decodings on $[[714,100]]$, $[[1428,184]]$ lifted-product codes under phenomenological noise model.}
   }\label{fig:lp}
\end{figure}

\emph{Lifted-Product Codes:} \fig{lp} presents the performance of sliding-window decoding on two lifted-product (LP) codes of parameters $[[714,100,\le 16]]$ and $[[1428,184, \le 24]]$.
The construction of these two codes are taken from Refs.~\cite{xu_constant-overhead_2023}. 
Compared to the HGP codes considered above, in LP codes we find that the advantage of overlapping decoding against non-overlapping decoding with the same $\width$ becomes more significant. This could be because the LP codes have larger distances and the error suppression would become more drastic even if the effective distance improves by a constant factor of the code distance.

\section{Conclusion and Outlook}\label{sec:outlook}
To summarize, we have studied the performance 
of sliding window decoder for qLDPC codes. 
Our hypothesis that this decoding strategy improves the effective code distance is confirmed by numerical simulations of the HGP and LP codes with varying window widths $\width$ and offset $\offset$. 
In particular we find that the lifetimes achieved with the overlapping $(\width,1)$-decoding strategy can be significantly larger than that with non-overlapping $(\width,\width)$-decoding strategy. 

While in this work we have only considered overlapping decoding with $F=1$, a larger offset $\offset$ might be desirable in order to correct more errors in one decoding cycle. In practice, the choices of $\width$ and $\offset$ will depend on the trade-off between logical error rate and decoding complexity.
If the desired logical error rate can be achieved with a small width, then it becomes unnecessary  to increase the width and needlessly increase the decoding complexity. It would also be beneficial to study parallel versions of overlapping decoding~\cite{tan_scalable_2022,skoric_parallel_2023} to ease the decoding complexity and improve speeds.

From a theoretical perspective, determining the minimum width $\width$ that preserves the code distance remains an open challenge. 
While for topological codes, it is widely accepted that $\width\ge d$ is necessary~\cite{dennis_topological_2002}, this may not necessarily be the case in general for qLDPC codes. Our conjecture is supported by the observation that the expected memory lifetime for the $[[625,25,8]]$ HGP code quickly increases with $\width$ in the beginning but saturates beyond $\width=5$, which is smaller than the code distance $d=8$. It would certainly be interesting if one could show existence of qLDPC code families for which $\width=o(d)$ is sufficient for distance-preserving decoding. 

Finally, it will be beneficial to 
simulate sliding window decoding under more realistic circuit-level noise models where gates used for syndrome extraction can be faulty. 
While, we leave these simulations for future work,
we expect that single-shot or non-overlapping decoding will degrade the effective distance more severely since  correlated errors may occur due to noise during gate operations composing the circuit.  
Thus, a slightly larger decoding window width may be required to maximize the expected memory lifetime. 

\section*{Acknowledgement}
The authors thank Yue Wu  for helpful discussion.
This material is based upon work supported by the U.S. Department of Energy, Office of Science, National Quantum Information Science Research Centers, Co-design Center for Quantum Advantage (C2QA) under contract number DE-SC0012704.

\bibliography{qLDPC}

\appendix

\section{Parity-check matrices for HGP codes}\label{app:HGP}

$[[625,25,8]]$:
\begin{eqnarray*}
A = \left[\begin{array}{c}
0 0 0 1 1 0 0 0 0 0 0 0 0 0 0 0 0 1 0 1\\
0 1 0 0 0 0 1 0 0 0 0 0 0 0 1 0 0 0 0 1\\
0 0 0 0 0 1 0 0 0 1 0 0 0 0 0 1 1 0 0 0\\
0 0 1 0 0 0 1 0 0 0 0 0 0 1 0 0 1 0 0 0\\ 
0 0 0 0 0 0 0 0 1 1 0 1 0 0 0 0 0 0 1 0\\ 
0 0 0 0 1 0 0 0 0 0 1 0 0 0 1 1 0 0 0 0\\ 
0 0 0 0 0 0 0 1 0 1 0 0 1 0 0 0 0 1 0 0\\ 
0 1 1 1 0 0 0 0 0 0 0 0 1 0 0 0 0 0 0 0\\ 
0 0 0 0 0 1 0 0 1 0 0 0 1 0 1 0 0 0 0 0\\ 
0 0 0 0 0 0 0 1 0 0 1 1 0 0 0 0 1 0 0 0\\ 
0 0 0 1 0 1 0 1 0 0 0 0 0 0 0 0 0 0 1 0\\ 
1 0 0 0 0 0 0 0 0 0 0 1 0 1 0 0 0 0 0 1\\ 
1 0 0 0 1 0 1 0 0 0 0 0 0 0 0 0 0 0 1 0\\ 
1 0 1 0 0 0 0 0 1 0 0 0 0 0 0 1 0 0 0 0\\ 
0 1 0 0 0 0 0 0 0 0 1 0 0 1 0 0 0 1 0 0
\end{array}\right]    
\end{eqnarray*}

$[[900,36,10]]$:
\begin{eqnarray*}
A = \left[
\begin{array}{c}
0 1 0 0 0 0 0 0 0 0 0 0 0 0 1 1 0 0 1 0 0 0 0 0\\ 
0 0 0 0 0 0 1 1 0 0 0 0 0 0 0 1 0 0 0 0 0 0 1 0\\ 
0 0 0 1 0 0 0 0 1 0 0 0 0 0 0 0 0 0 0 1 0 0 0 1\\ 
0 0 1 0 0 0 0 0 0 0 0 1 0 1 0 0 0 0 0 0 0 0 0 1\\ 
0 0 0 0 0 0 0 0 0 1 0 1 0 0 0 0 1 0 0 1 0 0 0 0\\ 
0 0 0 0 0 1 0 0 1 0 0 0 0 0 0 0 0 1 0 0 0 0 1 0\\ 
1 1 0 0 0 0 0 1 0 0 0 0 0 0 0 0 1 0 0 0 0 0 0 0\\ 
0 0 0 0 0 1 0 0 0 0 0 0 1 1 1 0 0 0 0 0 0 0 0 0\\ 
1 0 0 0 0 0 0 0 0 0 1 0 0 0 0 0 0 0 1 0 0 0 0 1\\ 
0 0 0 0 1 0 0 0 0 0 0 0 0 0 0 0 1 0 0 0 0 1 1 0\\ 
0 1 0 0 1 0 1 0 0 0 0 1 0 0 0 0 0 0 0 0 0 0 0 0\\ 
0 0 0 1 0 0 0 1 0 0 0 0 0 0 1 0 0 1 0 0 0 0 0 0\\ 
0 0 0 0 0 0 0 0 1 0 1 0 0 0 0 0 0 0 0 0 1 1 0 0\\ 
1 0 0 0 1 0 0 0 0 0 0 0 0 0 0 0 0 0 0 1 1 0 0 0\\ 
0 0 1 0 0 1 1 0 0 0 1 0 0 0 0 0 0 0 0 0 0 0 0 0\\ 
0 0 0 0 0 0 0 0 0 1 0 0 0 1 0 0 0 0 1 0 0 1 0 0\\ 
0 0 1 0 0 0 0 0 0 0 0 0 1 0 0 1 0 1 0 0 0 0 0 0\\
0 0 0 1 0 0 0 0 0 1 0 0 1 0 0 0 0 0 0 0 1 0 0 0
\end{array}\right]
\end{eqnarray*}

$[[1225,49]]$:
\begin{eqnarray*}
A = \left[
\begin{array}{c}
0 0 0 0 0 1 0 0 0 0 0 0 1 0 1 0 0 0 0 1 0 0 0 0 0 0 0 0\\
0 1 0 1 0 0 0 1 0 0 0 0 0 0 0 0 1 0 0 0 0 0 0 0 0 0 0 0\\
0 0 0 0 0 0 0 0 1 0 0 0 0 1 0 1 0 0 0 0 0 1 0 0 0 0 0 0\\
1 1 0 0 0 0 0 0 0 0 0 0 0 0 0 0 0 0 0 0 0 0 0 1 1 0 0 0\\ 
0 0 0 0 1 0 1 0 0 0 0 0 0 0 0 1 0 0 0 1 0 0 0 0 0 0 0 0\\ 
0 0 0 0 0 0 0 0 0 0 0 1 0 0 0 0 0 0 1 0 0 0 1 0 0 0 0 1\\ 
0 1 0 0 1 0 0 0 0 0 1 0 0 0 0 0 0 1 0 0 0 0 0 0 0 0 0 0\\ 
1 0 0 1 0 0 0 0 0 0 0 0 0 0 0 0 0 0 0 1 0 0 1 0 0 0 0 0\\ 
0 0 0 0 0 0 0 0 1 0 1 0 0 0 0 0 1 0 0 0 0 0 0 0 0 0 0 1\\ 
0 0 0 0 0 0 0 0 0 0 0 0 0 0 0 0 0 1 0 0 0 0 1 0 0 1 1 0\\ 
0 0 1 0 0 0 0 1 0 0 0 0 0 0 0 1 0 1 0 0 0 0 0 0 0 0 0 0\\ 
1 0 1 0 0 0 0 0 0 0 0 0 0 0 1 0 0 0 1 0 0 0 0 0 0 0 0 0\\ 
0 0 0 0 1 0 0 0 0 0 0 0 0 0 0 0 0 0 1 0 1 1 0 0 0 0 0 0\\ 
0 0 0 0 0 0 0 0 0 0 0 1 1 1 0 0 0 0 0 0 0 0 0 0 1 0 0 0\\ 
0 0 0 0 0 1 0 0 0 0 0 0 0 0 0 0 0 0 0 0 0 0 0 1 0 1 0 1\\ 
0 0 0 0 0 0 1 0 0 1 0 0 0 0 0 0 0 0 0 0 0 0 0 0 1 0 1 0\\ 
0 0 1 0 0 0 0 0 0 1 1 1 0 0 0 0 0 0 0 0 0 0 0 0 0 0 0 0\\ 
0 0 0 0 0 0 0 1 0 0 0 0 1 0 0 0 0 0 0 0 1 0 0 1 0 0 0 0\\ 
0 0 0 0 0 0 0 0 1 0 0 0 0 0 1 0 0 0 0 0 1 0 0 0 0 0 1 0\\ 
0 0 0 1 0 1 0 0 0 1 0 0 0 0 0 0 0 0 0 0 0 1 0 0 0 0 0 0\\ 
0 0 0 0 0 0 1 0 0 0 0 0 0 1 0 0 1 0 0 0 0 0 0 0 0 1 0 0
\end{array}
\right]  
\end{eqnarray*}

$[[1600,64]]$:
\begin{eqnarray*}
    A=\left[\begin{array}{c}
    0 1 0 0 0 0 0 0 0 0 0 0 0 0 0 0 0 0 0 0 0 0 0 0 0 1 1 0 1 0 0 0\\
0 0 0 0 0 0 0 0 0 0 0 0 0 0 0 1 0 0 0 0 0 0 0 0 1 0 0 0 0 0 1 1\\ 
0 0 0 0 0 0 1 0 0 1 0 0 0 0 0 0 0 0 0 0 0 0 0 0 0 0 1 1 0 0 0 0\\ 
1 0 0 0 0 0 0 0 0 0 0 0 0 0 0 0 0 0 0 1 1 0 0 0 0 0 0 0 0 1 0 0\\ 
0 0 0 1 0 0 0 0 0 0 0 0 1 0 0 0 0 0 0 0 0 1 0 0 0 0 0 0 0 0 0 1\\ 
0 0 1 1 1 0 0 0 0 0 0 0 0 0 0 0 1 0 0 0 0 0 0 0 0 0 0 0 0 0 0 0\\ 
1 0 0 0 0 0 0 0 0 0 0 0 0 1 0 0 0 0 1 0 0 0 0 0 0 1 0 0 0 0 0 0\\ 
1 0 0 0 1 0 1 0 0 0 0 0 0 0 0 1 0 0 0 0 0 0 0 0 0 0 0 0 0 0 0 0\\ 
0 0 0 0 0 0 0 1 0 0 1 0 0 1 0 0 0 0 0 0 0 1 0 0 0 0 0 0 0 0 0 0\\ 
0 0 0 0 0 0 0 1 0 0 0 0 0 0 0 0 1 0 0 0 0 0 0 1 0 0 1 0 0 0 0 0\\ 
0 1 0 0 0 1 0 0 0 0 0 0 0 0 1 1 0 0 0 0 0 0 0 0 0 0 0 0 0 0 0 0\\ 
0 0 0 0 1 0 0 0 0 0 0 0 0 0 0 0 0 1 0 0 0 0 1 0 0 0 0 0 1 0 0 0\\ 
0 0 0 0 0 0 0 0 1 0 0 0 0 0 0 0 0 0 1 0 0 0 0 0 0 0 0 1 0 0 1 0\\ 
0 0 0 0 0 1 0 1 1 0 0 0 0 0 0 0 0 1 0 0 0 0 0 0 0 0 0 0 0 0 0 0\\ 
0 1 1 0 0 0 0 0 0 1 0 0 0 0 0 0 0 0 0 0 1 0 0 0 0 0 0 0 0 0 0 0\\ 
0 0 1 0 0 0 0 0 0 0 1 0 0 0 1 0 0 0 0 0 0 0 0 0 0 0 0 0 0 0 1 0\\ 
0 0 0 0 0 1 0 0 0 1 0 0 1 1 0 0 0 0 0 0 0 0 0 0 0 0 0 0 0 0 0 0\\ 
0 0 0 0 0 0 0 0 0 0 0 0 0 0 0 0 0 0 1 0 0 0 1 1 0 0 0 0 0 0 0 1\\ 
0 0 0 0 0 0 0 0 0 0 0 1 0 0 1 0 0 0 0 0 0 0 0 1 0 0 0 0 0 1 0 0\\ 
0 0 0 0 0 0 1 0 0 0 1 0 0 0 0 0 0 0 0 1 0 0 1 0 0 0 0 0 0 0 0 0\\ 
0 0 0 0 0 0 0 0 0 0 0 1 0 0 0 0 1 0 0 0 0 0 0 0 1 1 0 0 0 0 0 0\\ 
0 0 0 1 0 0 0 0 0 0 0 0 0 0 0 0 0 1 0 0 0 0 0 0 0 0 0 1 0 1 0 0\\ 
0 0 0 0 0 0 0 0 0 0 0 0 1 0 0 0 0 0 0 1 0 0 0 0 1 0 0 0 1 0 0 0\\ 
0 0 0 0 0 0 0 0 1 0 0 1 0 0 0 0 0 0 0 0 1 1 0 0 0 0 0 0 0 0 0 0
    \end{array}\right]
\end{eqnarray*}

$[[2500,100]]$:
\begin{eqnarray*}
A = \left[\begin{array}{c}
0 0 0 0 0 0 0 0 0 0 0 0 0 1 0 0 0 0 0 0 0 1 0 0 0 0 0 0 0 1 0 0 0 0 0 0 0 1 0 0\\
1 0 0 1 0 0 0 0 0 0 0 0 0 0 0 0 0 0 0 0 0 0 0 1 0 0 0 0 0 0 0 0 0 0 0 0 0 0 0 1\\
  0 1 0 0 0 0 1 0 0 0 0 0 0 0 0 0 0 0 0 0 0 1 0 0 0 0 0 0 0 0 0 0 0 0 0 0 0 0 1 0\\
  0 0 0 0 1 0 0 0 0 0 0 0 0 0 0 0 0 0 0 0 0 0 0 0 0 0 1 1 0 0 0 0 0 0 1 0 0 0 0 0\\
  1 0 0 0 0 0 0 0 0 0 0 0 0 0 0 0 1 0 0 0 0 0 1 0 0 0 0 0 0 0 0 0 1 0 0 0 0 0 0 0\\
  0 0 0 0 0 0 0 0 0 0 0 0 0 1 0 0 0 0 0 0 0 0 0 1 0 0 0 0 0 0 1 0 0 0 0 1 0 0 0 0\\
  0 0 1 0 0 0 0 0 0 0 0 1 1 0 0 1 0 0 0 0 0 0 0 0 0 0 0 0 0 0 0 0 0 0 0 0 0 0 0 0\\
  0 0 0 0 0 0 0 1 0 0 0 1 0 0 0 0 1 0 0 1 0 0 0 0 0 0 0 0 0 0 0 0 0 0 0 0 0 0 0 0\\
 0 0 1 0 0 0 0 0 1 0 0 0 0 0 0 0 0 1 0 0 0 0 0 0 0 0 0 0 0 0 0 0 0 0 1 0 0 0 0 0\\
 0 0 0 0 0 0 0 0 1 0 1 0 0 0 0 0 0 0 0 0 0 0 0 0 0 0 0 0 0 1 0 1 0 0 0 0 0 0 0 0\\
 0 0 0 0 0 0 0 0 0 0 0 0 0 0 0 0 0 0 1 0 0 0 0 0 0 1 1 0 0 0 0 0 1 0 0 0 0 0 0 0\\
 0 0 0 1 0 0 0 0 0 0 0 0 0 0 0 0 0 0 1 0 0 0 0 0 0 0 0 0 0 0 0 0 0 0 0 0 1 1 0 0\\
 0 0 0 0 0 0 0 0 0 0 0 0 0 0 0 1 0 0 0 0 1 0 0 0 0 0 0 1 0 0 1 0 0 0 0 0 0 0 0 0\\
 0 0 0 0 0 1 0 0 0 0 1 0 0 0 1 0 0 0 0 0 0 0 0 0 0 0 0 0 0 0 0 0 0 0 0 1 0 0 0 0\\
 0 0 0 0 0 0 1 0 0 0 0 0 0 0 0 0 0 0 0 0 0 0 0 1 0 0 0 0 1 0 0 0 0 0 1 0 0 0 0 0\\
 0 0 0 0 0 0 0 0 0 0 0 0 1 1 0 0 0 0 0 0 0 0 0 0 1 0 0 0 0 0 0 0 1 0 0 0 0 0 0 0\\
 0 0 0 0 0 0 0 1 0 0 0 0 0 0 0 0 0 0 0 0 0 0 0 0 0 1 0 0 0 0 0 0 0 0 0 1 0 0 1 0\\
 0 0 1 0 0 0 0 0 0 0 0 0 0 0 0 0 0 0 0 0 0 0 0 0 0 1 0 0 0 0 0 0 0 1 0 0 0 0 0 1\\
 0 0 0 1 0 0 0 1 1 0 0 0 0 0 0 0 0 0 0 0 1 0 0 0 0 0 0 0 0 0 0 0 0 0 0 0 0 0 0 0\\
 0 1 0 0 1 0 0 0 0 0 0 0 0 0 0 0 0 0 0 1 0 0 0 0 0 0 0 0 0 0 0 0 0 0 0 0 0 0 0 1\\
 0 0 0 0 1 1 0 0 0 0 0 0 0 0 0 0 0 0 0 0 0 0 1 0 0 0 0 0 0 0 0 0 0 0 0 0 0 1 0 0\\
 0 0 0 0 0 0 0 0 0 0 0 0 0 0 1 0 1 1 0 0 0 1 0 0 0 0 0 0 0 0 0 0 0 0 0 0 0 0 0 0\\
 0 0 0 0 0 0 0 0 0 1 0 1 0 0 0 0 0 0 0 0 0 0 0 0 0 0 1 0 0 1 0 0 0 0 0 0 0 0 0 0\\
 1 0 0 0 0 0 0 0 0 0 0 0 0 0 0 0 0 0 0 0 0 0 0 0 0 0 0 1 0 0 0 1 0 0 0 0 0 0 1 0\\
 0 1 0 0 0 0 0 0 0 0 1 0 1 0 0 0 0 0 0 0 0 0 0 0 0 0 0 0 1 0 0 0 0 0 0 0 0 0 0 0\\
 0 0 0 0 0 1 1 0 0 0 0 0 0 0 0 1 0 0 1 0 0 0 0 0 0 0 0 0 0 0 0 0 0 0 0 0 0 0 0 0\\
 0 0 0 0 0 0 0 0 0 0 0 0 0 0 0 0 0 0 0 1 0 0 0 0 1 0 0 0 0 0 0 1 0 0 0 0 1 0 0 0\\
 0 0 0 0 0 0 0 0 0 1 0 0 0 0 0 0 0 0 0 0 0 0 1 0 0 0 0 0 1 0 0 0 0 1 0 0 0 0 0 0\\
 0 0 0 0 0 0 0 0 0 0 0 0 0 0 1 0 0 0 0 0 1 0 0 0 1 0 0 0 0 0 0 0 0 1 0 0 0 0 0 0\\
 0 0 0 0 0 0 0 0 0 1 0 0 0 0 0 0 0 1 0 0 0 0 0 0 0 0 0 0 0 0 1 0 0 0 0 0 1 0 0 0
\end{array}\right]
\end{eqnarray*}

\end{document}